\documentstyle[aps,twocolumn]{revtex}

\newcommand\inp{-\!\!\!\!\!\int}%PRINCIPAL VALUE
\newlength{\len}
\def\intp#1#2{{\renewcommand{\arraystretch}{0.4}%
                \settowidth\len{$\displaystyle\int$}%
                 \begin{array}{c}%
                 \makebox[0pt]{\hspace{ \len}$\scriptstyle{#2}$}\\
                 \mbox{${\displaystyle{-\!\!\!\!\!\!\int}}$}\\%
                 \makebox[0pt]{\hspace{- \len}$\scriptstyle{#1}$}
                 \end{array}%
                 }\! }
\begin{document}
\title{Universal Quantum Signatures of Chaos in
Ballistic Transport}
\author{R. A. Jalabert$^{(1)}$, J.-L. Pichard$^{(2)}$,
and C. W. J. Beenakker$^{(3)}$}
\address{$^{(1)}$Division de Physique Th\'eorique\cite{CNRS},
Institut de Physique Nucl\'eaire,
91406 Orsay Cedex, France}
\address{$^{(2)}$CEA, Service de Physique de l'Etat Condens\'e,
Centre d'\'Etudes de Saclay,\\
91191 Gif-sur-Yvette Cedex, France}
\address{$^{(3)}$Instituut-Lorentz, University of Leiden,
P.~O.~Box 9506, 2300 RA Leiden, The Netherlands\medskip\\
{\sf (Submitted to Europhysics Letters, March 16, 1994)}\medskip\\
\parbox{14cm}{\rm
The conductance of a ballistic quantum dot (having chaotic classical
dynamics and being coupled by ballistic point contacts to two electron
reservoirs) is computed on the single assumption that its scattering
matrix is a member of Dyson's circular ensemble. General formulas are
obtained for the mean and variance of transport properties in the
orthogonal ($\beta=1$), unitary ($\beta=2$), and symplectic ($\beta=4$)
symmetry class. Applications include universal conductance
fluctuations, weak localization, sub-Poissonian shot noise, and
normal-metal--superconductor junctions. The complete distribution
$P(g)$ of the conductance $g$ is computed for the case that the
coupling to the reservoirs occurs via two quantum point contacts with a
single transmitted channel. The result $P(g)\propto g^{-1+\beta/2}$ is
qualitatively different in the three symmetry classes.\smallskip
{\sf cond-mat/9403073}}}

\maketitle

\narrowtext

The search for signatures of chaotic behavior in quantum mechanical
systems \cite{Haake} has recently been extended to semiconductor
nanostructures known as ``quantum dots'' \cite{Mar92,ChaosRev}.
A quantum dot is
essentially a mesoscopic electron billiard, consisting of a ballistic
cavity connected by two small holes to two electron reservoirs. An
electron which is injected through one of the holes will either return
through the same hole, with probability $R$, or be transmitted through
the other hole, with probability $T$.  Classically, the uniform
(ergodic) exploration of the boundaries yields $T=R$, if the two holes
are of the same size and sufficiently small that direct transmission
(without boundary reflections) can be ignored.

For a {\em closed\/} quantum dot (without holes), it is well known that
one of the quantum signatures of its classically chaotic character
consists in the Wigner-Dyson distribution of the energy levels
\cite{Bohigas,Berry}. The Wigner-Dyson distribution was originally
derived by random-matrix theory (RMT), and is characterized by a
repulsion of nearby levels which depends only on the symmetry of the
Hamiltonian. The quantum analog of the ergodic exploration of the dot
boundaries by the classical trajectories consists in the Porter-Thomas
distribution of the eigenfunctions, as confirmed by numerical studies
of their amplitude distribution at the boundaries \cite{JalStoAl}.  The
quantum dot with holes is an {\em open}, rather than a closed system.
Just as the Wigner-Dyson distribution describes the Hamiltonian $H$ of
the closed system, Dyson's circular ensemble \cite{Dyson1} provides the
statistical properties of the scattering matrix $S$ of the open
system.  To have spectral or scattering properties given by the
universal RMT description for $H$ or $S$ can be actually regarded as a
precise definition of the somewhat vague concept of ``quantum chaos''.
To what extent a real ballistic cavity is close to this precise
universal limit is the subject of the theory of quantum billiards
\cite{Berry,Smilansky}.

Assuming this definition of a quantum chaotic system --- and this is
our only assumption --- we will calculate the statistics of the
transmission and reflection eigenvalues of the quantum dot, and hence
its transport properties. This allows us to determine the universal
quantum signatures of chaos in ballistic transport. Our investigation
was motivated by a remarkable calculation by P. A. Mello of the
variance of the conductance in the circular unitary ensemble
\cite{Mello-unpubl}. The approach presented below recovers his result
as a special case, and puts the quantum transport theory for a
ballistic chaotic billiard on the same footing as the established
theory for a disordered wire.

Dyson's circular ensemble characterizes a system where all scattering
processes are equally probable, subject to the constraints of current
conservation and time-reversal and spin-rotation symmetry. There exist
three symmetry classes: If a magnetic field $B$ is applied, $S$ is only
unitary ($\beta=2$, unitary ensemble); When $B=0$, $S$ is a unitary
symmetric matrix in the absence of spin-orbit scattering ($\beta=1$,
orthogonal ensemble) or otherwise a unitary self-dual quaternion matrix
($\beta=4$, symplectic ensemble). We assume the usual definition of
$S$, in terms of $N\!\times\!N$ submatrices describing how the incoming
modes are reflected or transmitted:
\begin{equation}
S = \left( \begin{array}{lr}
r	& \hspace{0.5cm} t'	\\
t	& \hspace{0.5cm} r'
\end{array} \right) \ .
\label{eq:Smat}
\end{equation}
$N$ is the number of transverse modes at the Fermi level in each of the
two leads connecting the dot to the reservoirs.  The probability
$P_{\beta}(dS)$ to find $S$ in a neighborhood $dS$ of some given $S$ is
\begin{equation}
P_{\beta}(dS)={ 1 \over V_{\beta}}\,\mu_{\beta}(dS),
\end{equation}
where $V_{\beta}=\int \mu_{\beta}(dS)$ is the total volume of the
$S$-matrix space and $\mu_{\beta}(dS)$ is the $\beta$-dependent measure
of the neighborhood $dS$ of $S$. In the original work of Dyson
\cite{Dyson1}, these measures are expressed in eigenvalue--eigenvector
coordinates. This is a suitable representation to obtain the
distribution of the scattering phase shifts, but is not very convenient
for a study of conduction through the quantum dot. A transport property
$A$ can generally be expressed as a linear statistic $ A=\sum_{n=1}^{N}
f(T_{n})$ on the transmission eigenvalues $T_{n}$. The $T_{n}$'s are
{\em not\/} eigenvalues of $S$ and are not in any simple way related to
the scattering phase shifts. Instead, $T_{n}$ is an eigenvalue of the
transmission matrix product $tt^{\dagger}$. The measures
$\mu_{\beta}(dS)$ have recently been calculated in the
transmission-eigenvalue representation \cite{JalabertPichard}. Since
this technical advance is at the basis of our analysis, we briefly
sketch the derivation for the orthogonal ensemble ($\beta=1$).

For $\beta=1$ the scattering matrix is unitary symmetric, so that it
can be represented in the form $S=YY^{\rm T}$, where $Y$ is unitary.
Note that this decomposition is not unique. An infinitesimal
neighborhood $dS$ of $S$ is given by $dS={\rm i}YdQY^{\rm T}$, with
$dQ$ a real symmetric matrix. It has been shown by Dyson \cite{Dyson1},
that if the matrix elements $dQ_{ij}$ vary through some small intervals
of lengths $d\mu_{ij}$, the measure $\mu_1$ equals $\mu_1(dS)= \prod_{i
\leq j} d\mu_{ij}$, independent of $Y$. We use this freedom to choose
$Y$ in the form
\begin{eqnarray}
Y={\cal U}{\cal O}{\cal I}&=&\left( \begin{array}{lr}
u	& \hspace{0.4cm} 0	\\
0	& \hspace{0.4cm} u'
\end{array} \right) \
\left( \begin{array}{cc}
({{\bf1}-\sqrt{\cal R} \over 2})^{1/2}	& \
-({{\bf1}+\sqrt{\cal R} \over 2})^{1/2}\\
({{\bf1}+\sqrt{\cal R} \over 2})^{1/2}
& \ \hphantom{-}({{\bf1}-\sqrt{\cal R} \over 2})^{1/2}
\end{array} \right)\nonumber\\
&&\hspace{3cm}\mbox{}\cdot \left( \begin{array}{cc}
{\bf 1}	& \hspace{0.4cm} 0	\\
0 & \hspace{0.4cm} {\rm i}\,{\bf 1}
\end{array} \right) ,
\end{eqnarray}
where ${\cal R}$ is an $N\!\times\!N$ diagonal matrix with elements
$R_{n}= 1-T_{n}$.  Since ${\rm i}dQ=dY^{\rm T}Y^*+Y^{\dagger}dY$, and
since $Y$ and $dY$ can be expressed in terms of the matrices ${\cal
U}$, ${\cal O}$, $\cal I$ and their neighborhoods $d\,{\cal U}$ and
$d{\cal O}$, one can easily get $dQ$ in  this parametrisation.  The
result is an expression for $\mu_1(dS)$ in terms of the measures
$\mu(d\,{\cal U})$ and $\mu(d{\cal R})$ associated with the matrices
${\cal U}$ and ${\cal R}$, times a Jacobian:
\begin{equation}
\mu_1(dS)=\prod_{i<j}\left| R_i - R_j \right| \prod_{i}
(1-R_{i})^{-1/2}
\mu(d{\cal R})\mu(d\,{\cal U}).
\end{equation}
Integration over the unitary matrix ${\cal U}$ gives the reflection
eigenvalue distribution $P(R_1,R_{2}, \ldots R_N)$ in the circular
orthogonal ensemble.  The calculations in the unitary
\cite{JalabertPichard} and symplectic \cite{Frahm} ensembles proceed
similarly.

The final result is conveniently written in terms of a new set of
variables $\lambda_{n}\in[0,\infty]$, related to the reflection and
transmission eigenvalues by $R_{n}\equiv\lambda_{n}/(1+\lambda_{n})$,
$T_{n}\equiv 1/(1+\lambda_{n})$. The distribution
$P(\lambda_{1},\lambda_{2},\ldots\lambda_{N})$ of the
$\lambda$-variables takes the form of a Gibbs distribution,
\begin{mathletters}
\label{Gibbs}
\begin{eqnarray}
&&P(\{\lambda_{n}\}) = Z^{-1}\exp[- \beta {\cal H}(\{\lambda_{n}\})],
\label{eq:Jacobian}\\
&&{\cal H}(\{\lambda_{n}\}) = - \sum_{i<j} \ln{\left|\lambda_{i} -
\lambda_{j} \right|} + \sum_{i} V_{\beta} (\lambda_{i}),
\label{eq:Hamiltonian}\\
&&V_{\beta}(\lambda) = \left( N + \frac{2 - \beta}{2\beta} \right) \
\ln{(1+\lambda)},
\label{eq:potential}
\end{eqnarray}
\end{mathletters}%
where $Z$ is a normalization constant. The symmetry parameter
$\beta\in\{1, 2, 4\}$ plays the role of an inverse temperature. The
fictitious ``Hamiltonian'' ${\cal H}$ consists of a logarithmic pairwise
interaction plus a one-body potential $V_{\beta}(\lambda)$. This
potential is symmetry-independent to order $N$, while the term of order
$N^{0}$ depends on $\beta$.

Remarkably, the distribution (\ref{Gibbs}) is identical to the global
maximum-entropy ansatz for the transfer matrix of a diffusive conductor
\cite{StoneMMP} --- except for the one-body potential, which is
different: The potential $V_{\rm d}(\lambda)$ for a disordered wire of
length $L$ and mean free path $l$ is given by \cite{StoneMMP}
\begin{equation}
V_{\rm d}(\lambda)=(Nl/L)\ln^{2}(\sqrt{\lambda}+\sqrt{1+\lambda})
+{\cal O}(N^{0}).
\end{equation}
The potential (\ref{eq:potential}), in contrast, contains no
microscopic parameters and increases more slowly with $\lambda$. In the
case of a disordered wire, it is known \cite{BeenakkerRejaei} that the
logarithmic repulsion $-\ln|\lambda_{i}-\lambda_{j}|$ is only
rigorously valid for the weakly reflected scattering channels
($\lambda_{i},\lambda_{j}\ll 1$). In the ballistic chaotic dot, the
logarithmic repulsion which we have found is a direct consequence of
the basic assumption that the scattering matrix belongs to the circular
ensemble. It remains to be seen whether this assumption breaks down in
some range of $\lambda$. It seems that this problem is unrelated to the
breakdown of the logarithmic repulsion in the statistics of the energy
levels of a chaotic system \cite{AltShk,JaPiBe1}, which occurs on
energy scales greater than the Thouless energy $E_{\rm c}$. This energy
scale has no obvious counterpart for the transmission eigenvalues.

We consider transport properties of the form
$A=\sum_{n=1}^{N}a(\lambda_{n})$. To calculate the expectation value
$\langle A\rangle=\int_0^{\infty} a(\lambda) \rho (\lambda) d\lambda$
for a ballistic chaotic system, we need the density $\rho (\lambda)$ of
the $\lambda$'s in the circular ensemble. For this purpose, we use
Dyson's large-$N$ expansion \cite{Dyson2}
\begin{equation}
\intp{0}{\infty}{\rho(\lambda')\over{\lambda-\lambda'}}\,d\lambda'
+{{\beta-2}\over {2\beta}} {d \over {d \lambda}} \ln \rho(\lambda) =
{d \over {d \lambda}} V_{\beta}(\lambda),
\label{eq:meanfield}
\end{equation}
where $\inp$ denotes the principal value.  We decompose
$\rho=\rho_{N}+\delta\rho$ into a contribution $\rho_N$ of order $N$
(giving the ``Boltzmann conductance'') and a symmetry-dependent
correction $\delta \rho$ of order $N^0$ (responsible for the
``weak-localization effect''). From Eqs.\ (\ref{Gibbs}) and
(\ref{eq:meanfield}) we find, order by order,\footnote{
The result (\protect\ref{eq:cor}) for the ${\cal O}(N^{0})$ correction
$\delta\rho$ holds only for $\lambda\ll N^{2/3}$, because for larger
$\lambda$'s the ${\cal O}(N)$ contribution $\rho_{N}\approx
N\lambda^{-3/2}$ no longer dominates the density and the large-$N$
expansion fails. The large-$\lambda$ tail ensures that
$\int_{0}^{\infty}\delta\rho(\lambda)\,d\lambda=0$, but is irrelevant
for the conductance of the quantum dot. [The range $\lambda\gtrsim
N^{2/3}\gg 1$ gives a contribution to $g$ of order $N^{-2/3}$, which
can be neglected relative to the contribution of order 1 which
is retained.]}
\begin{eqnarray}
\intp{0}{\infty}{ \rho_N (\lambda') \over {\lambda -\lambda'}}
\,d\lambda'= {N \over 1+\lambda}&\Rightarrow&
\rho_N(\lambda) = \frac{N}{\pi (1+\lambda) \sqrt{\lambda}},
\label{eq:rhon}\\
\intp{0}{\infty} {\delta \rho(\lambda') \over {\lambda-\lambda'}}
\,d\lambda'={\frac{\beta-2}{4\beta}}\,{ 1 \over \lambda}&\Rightarrow&
\delta \rho (\lambda)={\frac{\beta-2}{4\beta}} \delta_{+}(\lambda),
\label{eq:cor}
\end{eqnarray}
where the one-sided delta-function satisfies $\int_0^{\infty}
\delta_{+}(\lambda)\,d\lambda=1$.

It is interesting to compare this eigenvalue density for a ballistic
chaotic cavity to the density $\rho^{({\rm d})}$ for a disordered wire.
To order $N$ and for $L\gg l$ one has \cite{Mello}
\begin{equation}
\rho^{({\rm d})}_N (\lambda)= {Nl \over 2L}
{1 \over \sqrt{\lambda(1+\lambda)}},\;{\rm for}\;
\lambda<\lambda_{\rm c}\simeq\case{1}{4}\exp(2L/l).
\label{rhowire}
\end{equation}
The density goes to zero abruptly near a cutoff $\lambda_{\rm c}\gg 1$,
in such a way that $\int_{0}^{\infty}\rho^{({\rm
d})}_{N}(\lambda)\,d\lambda=N$. The transmission eigenvalue density
$\rho_{N}(T)=\rho_N(\lambda)|d\lambda/dT|$ (with $T=(1+\lambda)^{-1}$)
is {\em bimodal\/}, with a peak near unit and near zero transmission.
This is a familiar result for a diffusive conductor, but was not
previously established for a chaotic dot.

The term of order $N^0$, which yields the weak-localization
(anti-localization) corrections for a disordered wire when $\beta\neq
2$, is given by \cite{Carlo1994}:
\begin{eqnarray}
\delta \rho^{({\rm d})}(\lambda)&=&\frac{\beta-2}{2\beta}\left[
\delta_{+}(\lambda)+(\lambda+\lambda^{2})^{-1/2}\right.\nonumber\\
&&\left.\mbox{}\times\biglb(
4\ln^{2}[\sqrt{\lambda}+\sqrt{1+\lambda}]+\pi^{2}\bigrb)^{-1}\right],
\label{weaklocawire}
\end{eqnarray}
which is not as strongly peaked near $\lambda=0$ as the delta-function
result (\ref{eq:cor}) for a chaotic dot.

We now use Eqs.\ (\ref{eq:rhon}) and (\ref{eq:cor}) to calculate the
expectation value $\langle T\rangle$ of the total transmission
probability $T={\rm Tr}\,tt^{\dagger}=\sum_{n}(1+\lambda_{n})^{-1}$.
According to the Landauer formula, $T$ equals the conductance $g$
(measured in units of $2e^2/h$ where the factor 2 comes from spin or
Kramers degeneracy). The result is
\begin{equation}
\langle T\rangle=\frac{1}{2}N+\delta T,\;\;
\delta T=\frac{\beta-2}{4\beta}.
\end{equation}
For $\beta=2$, one finds $\langle T\rangle=\frac{1}{2}N=\langle
R\rangle$ (where $\langle R\rangle =N-\langle T\rangle$ is the total
reflection probability). This is the quantum analog of what we expect
from the ``ergodic'' exploration of the dot boundaries by the classical
trajectories.  Quantum interference then breaks the equality $\langle
T\rangle=\langle R\rangle$ by an amount $\delta T$, due to weak
localization ($\beta =1$) or anti-localization ($\beta=4$). The value
$\delta T=-\frac{1}{4}$ for $\beta=1$ is in agreement with the results
of Ref.\ \cite{Iid90} and demonstrates the point raised in
Ref.\ \cite{BarJalSto}: Weak localization is not only given by coherent
backscattering, but has an off-diagonal (in mode index and classical
trajectory labels) component. In the same way one can compute the
average of any other linear statistic. We give two examples to
illustrate the generality of our approach.

The first example is the shot-noise power $P$ which is given
by \cite{Buttiker}
$P=P_{0}{\rm Tr}\,tt^{\dagger}({\bf 1}-tt^{\dagger})$, with
$P_{0}=(2e^{2}/h)2eU$ ($U$ is the applied voltage). In this case
$a(\lambda)=P_{0}\lambda(1+\lambda)^{-2}$. Since
$\delta\rho(\lambda)a(\lambda)\equiv 0$ for any $\beta$, there is {\em
no\/} weak-localization correction for the shot noise of a chaotic dot
--- in contrast to a diffusive conductor, where a
weak-localization effect does exist \cite{Jong}. Integration of
$\rho_{N}(\lambda)a(\lambda)$ gives the average shot-noise power
\begin{equation}
\langle P\rangle =\frac{1}{8}NP_{0}=\frac{1}{4}P_{\rm Poisson},
\end{equation}
which is four times smaller than the Poisson noise $P_{\rm
Poisson}=gP_{0}=2eI$ associated with a current $I$ of uncorrelated
electrons. The $\frac{1}{4}$ reduction in a chaotic dot is to be
compared with the $\frac{1}{3}$ reduction of shot noise in a diffusive
conductor \cite{CarloMarkus}.

The second example is the conductance $G_{\rm NS}$ of the dot if one of
the two attached reservoirs is a superconductor. This case corresponds
to \cite{Beenakker1} $a(\lambda)=(4e^{2}/h)(1+2\lambda)^{-2}$ if
$\beta=1, 4$ ($G_{\rm NS}$ is not a linear statistic for $\beta = 2$).
Again, we find a noticeable difference between the diffusive disordered
wire and the ballistic chaotic dot. In the disordered case, the
conductance $G_{\rm N}$ in the normal state is unchanged if one of the
reservoirs becomes superconducting ($\langle G_{\rm N}\rangle=\langle
G_{\rm NS}\rangle$, up to a weak-localization correction of order
$N^{0}$). In the ballistic chaotic case, Eq.\ (\ref{eq:rhon}) yields
(to order $N$):
\begin{equation}
\langle G_{\rm NS}\rangle={2e^2 \over h}(2-\sqrt{2})N,
\end{equation}
which differs to order $N$ from the result $\langle G_{\rm
N}\rangle=(2e^{2}/h)(N/2)$ in the normal state. Another example, which
we do not discuss here but which can be treated in just the same way
\cite{Beenakker1}, is the Josephson effect if both reservoirs are
superconducting.

So far we have focused on the expectation values in the circular
ensemble. Fluctuations around the average in this ensemble can be
computed using the general formulas of Ref.\ \cite{Beenakker1}, which
hold for any ensemble with a logarithmic interaction (regardless of the
form of the one-body potential). The variance in the large-$N$ limit is
given by
\begin{eqnarray}
{\rm Var}\,A&=&-\frac{1}{\beta}\,\frac{1}{\pi^{2}}\int_{0}^{\infty}
\!\!d\lambda\int_{0}^{\infty}
\!\!d\lambda'\left(\frac{da(\lambda)}{d\lambda}\right)
\left(\frac{da(\lambda')}{d\lambda'}\right)\nonumber\\
&&\hspace{3cm}\mbox{}\times\ln
\left|\frac{\surd\lambda-\surd\lambda'}
{\surd\lambda+\surd\lambda'}\right|.\label{variance}
\end{eqnarray}
Eq.\ (\ref{variance}) can be used to compute the analogue of the
``Universal Conductance Fluctuations'' (UCF) in a ballistic chaotic
cavity. These fluctuations can be induced by a change of the Fermi
energy, of the applied magnetic field, or of the boundary spin-orbit
scattering, as well as by a slight deformation of the shape of the
dot.  One obtains, for example, ${\rm Var}\,g=1/8\beta$, ${\rm
Var}\,P/P_{0}=1/64\beta$ for the fluctuations in the conductance and
shot noise, respectively. The $1/\beta$-dependence is the same as for a
disordered wire, but the numerical coefficients are somewhat different
due to the difference in interaction potential \cite{BeenakkerRejaei}.

The results for mean and variance given above require $N\gg 1$. The
opposite regime $N=1$ is also of interest. This would apply to a
semiconductor quantum dot which is coupled to the reservoirs by two
quantum point contacts with a quantized conductance of $2e^{2}/h$. The
probability distribution (\ref{Gibbs}) reduces for $N=1$ to
$P(\lambda)=\frac{1}{2}\beta(1+\lambda)^{-1-\beta/2}$. This implies for
the (dimensionless) conductance $g=(1+\lambda)^{-1}$ the distribution
\begin{equation}
P(g)=\case{1}{2}\beta\,g^{-1+\beta/2},\;\;0\leq g\leq 1.
\label{Pofg}
\end{equation}
This is a remarkable result: In the presence of magnetic field
($\beta=2$), any value of the conductance between $0$ and $2e^{2}/h$ is
equally probable. In non-zero field it is more probable to find a small
than a large conductance, provided that the boundary scattering
preserves spin-rotation symmetry ($\beta=1$). In the presence of
spin-orbit scattering at the boundary ($\beta=4$), however, a large
conductance is more probable than a small one. To observe this
qualitatively different behavior presents a challenge for
experimentalists.

In summary, we have calculated the distribution of the transmission and
reflection eigenvalues characterizing Dyson's circular ensemble.
Relying on a definition of ``quantum chaos'' based on the applicability
of this ensemble to describe scattering in a ballistic chaotic cavity,
we have extracted from the joint probability distribution (\ref{Gibbs})
the expectation values and the fluctuations of arbitrary linear
statistics related to quantum transport. We have mainly stressed the
differences between ballistic chaotic dots and disordered wires.
Another important point, which we underline in conclusion, consists in
the qualitative differences existing between a fully chaotic dot and a
dot where the classical dynamics is integrable and for which larger
quantum fluctuations are generally expected \cite{Mar92,ChaosRev}.

This work was supported in part by EEC, Contract No. SCC--CT90--0020,
and by the Dutch Scienc Foundation NWO/FOM.

{\em Note added:} Upon completion of this manuscript we received a
preprint by Baranger and Mello, in which some of our
results are obtained by a different method.

\end{document}